\documentclass{JHEP}
\usepackage{epsfig}
\usepackage{graphicx}%
\def\be{\begin{equation}}
\def\ee{\end{equation}}
\def\bea{\begin{eqnarray}}
\def\eea{\end{eqnarray}}
 
\def\lesssim{\mathrel{\hbox{\rlap{\hbox{\lower4pt\hbox{$\sim$}}}\hbox{$<$}}}}
\def\gtrsim{\mathrel{\hbox{\rlap{\hbox{\lower4pt\hbox{$\sim$}}}\hbox{$>$}}}}

\title{Inflationary Theory and Alternative Cosmology}

\author{Lev Kofman\\
    CITA, University of
Toronto, 60 St. George Street, Toronto, ON M5S 3H8, Canada\\ E-mail: 
\email{kofman@cita.utoronto.ca}}
\author{Andrei Linde\\
    Department of Physics, Stanford University, Stanford, CA 94305,
USA\\
    E-mail: \email{alinde@stanford.edu}, ~~ 
http://physics.stanford.edu/linde}

\author{V. Mukhanov\\
  Sektion Physik, Ludvig Maximillian University,
37 Teresienstr., Munich, Germany, ~
and Princeton University,
Princeton, NJ 08544, USA\\ E-mail: 
\email{ mukhanov@theorie.physik.uni-muenchen.de }}

\received{\today}       
\preprint{CITA-2002-15\\ SU-ITP-02/26\\ \hepth{0206088}\\ June 11, 2002}

\abstract{Recently Hollands and Wald argued that inflation does not
solve any of the major cosmological problems. We explain why we
disagree with their arguments. They also proposed a new speculative mechanism
of generation of density perturbations. We show that in their scenario
the inhomogeneities responsible for the large scale structure observed
today were generated at an epoch when the energy density of the hot
universe was $10^{95}$ times greater than the Planck density. The only
way to avoid this problem is to assume that there was a stage of
inflation in the early universe.}

\begin{document}

\section{Introduction}
During the last 20 years inflationary theory \cite{Guth,New,Chaot} has
evolved from a problematic hypothesis to an almost universally
accepted cosmological paradigm \cite{book}. It solves many fundamental
cosmological problems and makes several predictions that agree very
well with the observational data \cite{Bond}. Despite this fact (or
maybe because of it) it has become popular to propose various
alternatives to inflation.

In this paper we will consider one such alternative suggested recently
by Hollands and Wald \cite{alt}. The authors admit that their model
does not solve or even address the homogeneity, isotropy, flatness,
horizon and entropy problems, but they claim that inflation does not
do so either. We will examine their claim and explain why we disagree
with it. We will use this discussion as an opportunity to emphasize
some properties of inflation that may not be widely known.

 What Hollands and Wald's model does attempt to explain is the
generation of density perturbations with a flat spectrum. However, one
cannot justify this mechanism using the standard methods of quantum
field theory. Moreover, in this scenario density perturbations on the
scale of the present horizon were generated at a time when the energy
density of the hot universe was $10^{95}$ times greater than the
Planck density. Since nobody knows how to make any calculations at
such densities, one must be hard pressed to consider this an
alternative to the inflationary mechanism of generation of density
perturbations. We will explain that the origin of this problem is
directly related to the absence of inflation in the model proposed in
\cite{alt}.

\section{Fluctuations in the model with fundamental scale without inflation}\label{alt1}

We will begin our discussion of the paper by Hollands and Wald
\cite{alt} by describing their proposed mechanism for the generation
of cosmological fluctuations without inflation.

Consider for definiteness radiation-dominated cosmological expansion
that begins at a singularity at $t=0$, with a scale factor
$a(t)=\sqrt{t}$.  According to \cite{alt}, one should introduce a new
``fundamental scale'' $l_0 \sim 10^{-5} M_p^{-1} \sim 10^{-28}$
cm.\footnote{The authors associate it with the GUT scale, even though
the GUT length scale is 3 orders of magnitude smaller. In fact, $l_0$
is exactly the scale corresponding to the inverse mass of the inflaton
field in chaotic inflation \cite{Chaot}.}  It is assumed in \cite{alt}
that quantum fluctuations evolve according to QFT only in the regime
where their physical wavelengths exceed the fundamental length,
$\lambda(t) \geq l_0$, where initially $l_0$ is much greater than the
Hubble radius $H^{-1}$.

In inflationary theory quantum fluctuations oscillate until their
wavelength reaches $H^{-1}$. After that they freeze with amplitude
$\delta\phi \sim {H\over 2\pi}$. This amplitude can be calculated
using standard quantum field theory methods \cite{fluct} and then it
can be used to calculate the amplitude of inflationary density
perturbations \cite{Mukh}.

But fluctuations with wavelength greater than $H^{-1}$ do not
oscillate even if the universe is not inflationary.  As a result, from
the very beginning of their QFT phase, quantum fluctuations in the
scenario proposed in \cite{alt} are frozen with almost constant
amplitude, which is supposed to be $\delta\phi \sim {l_0}^{-1}$. This
last assumption does not follow from quantum field theory in curved
space. This could be a reasonable assumption if $l_0^{-1}$ were the
largest parameter of dimension of mass, but the authors use this
assumption in the situation when  $H \gg l_0^{-1}$, i.e. when the
size of the horizon $H^{-1}$ is much smaller than the fundamental
length $l_0$. We will return to this important point later. However,
if this assumption is correct, the spectrum of perturbations generated
by this mechanism for $l_0 \sim 10^{-5} M_p$ can be comparable to the
spectrum of standard inflationary perturbations.

  One can estimate the time $t_*$ when the fluctuations
corresponding to the present cosmological horizon $10^{28}$ cm were
frozen.  If the scale factor at present is unity, then
\begin{equation}\label{scale}
a(t_*) =    \frac{l_0}{10^{28} cm}  \simeq 10^{-56} \ .
\end{equation}
The density of radiation today is $\rho_r \simeq 10^{-35}$g/cm$^3 \sim
10^{-129} M_p^4$. This density at the time $t_*$ was greater by a
factor of $a^{-4}(t_*) \sim 10^{224}$.  Therefore at the moment $t_*$
the density of radiation was
\begin{equation}\label{density}
 \rho_r(t_*) \simeq    10^{95} M_p^4 \ .
\end{equation}
Here $M_p^4 \simeq 10^{94}$ g/cm$^3$ is the Planck density, $M_p =
G^{-1/2} \sim 10^{19}$ GeV.

But this means that all assumptions concerning the generation of
perturbations made in \cite{alt} are completely unreliable.  
Indeed, quantum fluctuations of the curvature of space and time at
densities $95$ orders of magnitude greater than the Planck density are
so large that any discussion of the universe in terms of classical
space-time becomes impossible.

One can easily verify that this problem appears not only in a
radiation dominated universe with $p = \rho/3$, but for all equations
of state $p = \omega \rho$ with $0<\omega< 1$ considered in
\cite{alt}.  Indeed, let us assume that physical laws up to the
electroweak scale $T_{_{\rm EW}} \sim 10^2$ GeV are well known, so
the universe was radiation dominated when it had electroweak scale
energy density. We will also assume that before the electroweak stage
the equation of state was $p = \omega \rho$.

At the electroweak stage the scale factor $a_{_{\rm EW}}$ was smaller
than at the present time by a factor of $2.7 K^o/T_{_{\rm EW}}\sim
10^{-15}$, and the density was $\rho_{_{\rm EW}} =O(T^4_{_{\rm EW}})
\sim 10^{-68} M_p^4$. The density perturbations on the scale of the
present horizon were generated at an epoch when the scale factor was
$a(t_*) = 10^{-56} = 10^{41} a_{_{\rm EW}}$, and the density was
\begin{equation}\label{density2}
 \rho_r(t_*) \simeq \rho_{_{\rm EW}} \left({a_{_{\rm EW}}\over a(t_*)}\right)^{3(1+\omega)} =  10^{-68} M_p^4 \times 10^{41(3(1+\omega))}   \ .
\end{equation}
This corresponds to a density smaller than $M_P^4$ only if $\omega <
-{55\over 123} \sim -0.45$. This means that $p < -0.45 \rho$ and $\rho
+ 3p <-0.3 \rho$, which corresponds to an accelerating (=inflating)
universe with negative pressure. If the standard regime $p = \omega
\rho$ with $0<\omega< 1$ took place all the way up to the GUT scale,
one can show that $\rho_r(t_*) < M_p^4$ only if $p \approx -\rho$
above the GUT scale. In other words, in order to avoid speculations
about super-Planckian densities in the mechanism of generation of
density perturbations proposed in \cite{alt} one must have a stage of
inflation in the early universe. Thus, the mechanism of \cite{alt}
does not offer any alternative to inflation because it needs inflation
for its own consistency.

The main difference between the models with ``normal'' equations of
state $0<\omega< 1$ considered in \cite{alt} and inflationary
cosmology is the following.  In the simplest versions of
inflationary cosmology the fluctuations responsible for the observed
CMB anisotropy also freeze on a length scale $\sim 10^{-5}M_p^{-1}$,
and then their wavelengths grow until they reach $10^{28}$ cm.  
However, {\it the energy density of the universe does not change much
during the first 60 e-folds of this growth}. That is why if one takes
perturbations on the scale $10^{28}$ cm and follows their evolution
back in time, one finds that they were produced during the stage of
inflation when the energy density was many orders of magnitude {\it
smaller}\, than the Planck density. This is one of the many amazing
features of inflationary cosmology. The attempt to reproduce
inflationary results \cite{alt} fails exactly because the authors were
trying to achieve them without using an early stage of inflation.
Paradoxically, evaluation of the ``alternative to inflation'' proposed
in \cite{alt} provides an additional argument in favour of inflation.

In fact, the situation with the mechanism proposed in \cite{alt} is
even more problematic. Consider again a radiation dominated universe
at the time $t_*$. The Hubble constant at that time is $10^{52}$ times
greater than $l_0^{-1}$ and the temperature is $10^{29}$ times greater
than $l_0^{-1}$. But how could it be possible that the wavelength of
the high-energy particles $T^{-1}$ is $29$ orders of magnitude smaller
than the ``elementary length'' $l_0$, and the size of horizon is $52$
orders of magnitude smaller than the ``elementary length'' $l_0$? This
contradicts the basic assumptions of \cite{alt}. And even if this were
possible, then the standard quantum field theory considerations would
suggest that at the time $t_*$ the amplitude of fluctuations would be
determined not by $l_0$, but by the greatest of these dimensional
parameters, i.e. it would be expected to be $10^{52}$ times greater
than the amplitude postulated in \cite{alt}.

Note that this problem persists even if one would attempt to apply the prescription
by Hollands and Wald to the generation of density perturbations during
inflation, where the energy density remains below the Planck
density. Indeed, the basic idea of this prescription is that the
wavelength {\it and the amplitude} of perturbations generated at $H
\gg l_0$ should be determined only by $l_0$. According to inflationary
theory, the amplitude of quantum fluctuations generated during
inflation is determined by the Hawking temperature in de Sitter space,
$\delta\phi \sim T_H ={H\over 2\pi}$.  We have no idea how one could
justify an assumption of \cite{alt} that the amplitude of
perturbations produced during inflation should be $\sim l_0^{-1}$,
which is much smaller than the Hawking temperature during
inflation. We believe that it is inconsistent to assume that the size
of the horizon in de Sitter space $H^{-1}$ is much smaller than the
``elementary length'' $l_0$.

\section{Initial conditions for inflation}

A significant part of \cite{alt} was devoted to a discussion of the
problem of initial conditions for inflation. In this section we will
describe the simplest model of chaotic inflation and consider the
issue of initial conditions. Then we will compare our analysis and the
arguments of \cite{alt}.

\subsection{Initial conditions for chaotic inflation}\label{cl}

Consider the simplest model of a scalar field $\phi$ with a mass $m$
and with the potential energy density $V(\phi) = {m^2\over 2}
\phi^2$. (One may also add a cosmological constant $V_0
\sim 10^{-123} M_p^4$ to describe the present stage of acceleration of
the universe.)  Since this function has a minimum at $\phi = 0$, one
may expect that the scalar field $\phi$ should oscillate near this
minimum. This is indeed the case if the universe does not expand, in
which case the equation of motion for the scalar field coincides with
the equation for a harmonic oscillator, $\ddot\phi = -m^2\phi$.

However, because of the expansion of the universe with Hubble constant
$H = \dot a/a $, an additional term $3H\dot\phi$ appears in the
harmonic oscillator equation:
\begin{equation}\label{1}
 \ddot\phi + 3H\dot\phi = -m^2\phi \ .
\end{equation}
The term $3H\dot\phi$ can be interpreted as a friction term.  The
Einstein equation for a homogeneous universe dominated by a scalar
field $\phi$ looks as follows:
\begin{equation}\label{2}
H^2 +{k\over a^2} ={8\pi\over 6}\, \left(\dot \phi
^2+m^2 \phi^2  \right) \ .
\end{equation}
Here $k = -1, 0, 1$ for an open, flat or closed universe
respectively. For simplicity, we work in units $M_p^{-2} = G = 1$.

If the scalar field $\phi$ initially was large, the Hubble parameter
$H$ was large too, according to the second equation. This means that
the friction term $3H\dot\phi$ was very large, and therefore the
scalar field was moving very slowly.  At this stage the energy density
of the scalar field remained almost constant, and the expansion of the
universe continued with a much greater speed than in the old
cosmological theory. Due to the rapid growth of the scale of the
universe and a slow motion of the field $\phi$, soon after the
beginning of this regime one has $\ddot\phi \ll 3H\dot\phi$, $H^2 \gg
{k\over a^2}$, $ \dot \phi ^2\ll m^2\phi^2$, so the system of
equations can be simplified:
\begin{equation}\label{E04}
H= {\dot a \over a}   =2m|\phi| \sqrt {\pi
\over 3}\ , ~~~~~~  \dot\phi = \pm {m\over 2  \sqrt{3\pi}}     .
\end{equation}
The first equation shows that if the field $\phi$ changes slowly, the
size of the universe in this regime grows approximately as $e^{Ht}$,
where $H = 2m|\phi| \sqrt {\pi \over 3}$. This is the stage of
inflation, which ends when the field $\phi$ becomes much smaller than
$M_p=1$.  A universe initially filled with a field $\phi = \phi_0
\gg 1$ will experience a long stage of inflation and grow
exponentially large.

 The issue of initial conditions in this scenario has been analysed
using various methods including Euclidean quantum gravity, the
stochastic approach to inflation, etc., see e.g. \cite{book}. Here we
would like to take a simple intuitive approach.

Consider, for definiteness, a closed universe of initial size $l \sim
1$ (in Planck units) that emerges from the space-time foam, or from a
singularity, or from `nothing,' in a state with Planck density $\rho
\sim 1$. Only starting from this moment, i.e. at $\rho \lesssim 1$,
can we describe this domain as a {\it classical} universe.  Thus, at
this initial moment the sum of the kinetic energy density, gradient
energy density, and potential energy density is of order unity:\,
${1\over 2} \dot\phi^2 + {1\over 2} (\partial_i\phi)^2 +V(\phi) \sim
1$.

It is important to understand that in this model there are no {\it a
priori} constraints on the initial value of the scalar field in this
domain, except for the constraint ${1\over 2} \dot\phi^2 + {1\over 2}
(\partial_i\phi)^2 +V(\phi) \sim 1$.  Indeed, let us consider for a
moment a theory with $V(\phi) = const$. This theory is invariant under
the shift $\phi\to \phi + c$. Therefore, in such a theory {\it all}
initial values of the homogeneous component of the scalar field $\phi$
are equally probable. Contrary to some assertions in the literature,
quantum gravity corrections do not lead to any constraints of the type
$\phi <1$ \cite{book}. Such corrections may affect the potential if, e.g., one tries to incorporate this model into N=1 supergravity.  But this is a separate issue of model building rather than the issue of initial conditions for our simple model. The only constraint on the average
amplitude of the field appears if the effective potential is not
constant, but grows and becomes greater than the Planck density at
$\phi > \phi_p$, where $V(\phi_p) = 1$. This constraint implies that
$\phi \lesssim \phi_p$, but it does not give any reason to expect that
$\phi \ll \phi_p$. This suggests that the typical initial value
$\phi_0$ of the field $\phi$ in such a theory is $\phi_0 \sim \phi_p$.
Thus, we expect that typical initial conditions correspond to ${1\over
2} \dot\phi^2 \sim {1\over 2} (\partial_i\phi)^2\sim V(\phi) = O(1)$.
If by any chance ${1\over 2} \dot\phi^2 + {1\over 2}
(\partial_i\phi)^2 \lesssim V(\phi)$ in the domain under
consideration, then inflation begins, and within a Planck time the
terms ${1\over 2} \dot\phi^2$ and ${1\over 2} (\partial_i\phi)^2$
become much smaller than $V(\phi)$, which ensures the continuation of
inflation. The probability of such an event may be equal to $1/2$, or
$10^{-1}$, or maybe even $10^{-2}$,  but there's no obvious reason
why it should be exponentially suppressed. Moreover, the total
lifetime of a non-inflationary universe with ${1\over 2} \dot\phi^2 +
{1\over 2} (\partial_i\phi)^2 > V(\phi)$ is $O(1)$ in Planck units,
i.e. $10^{-43}$ seconds. Such universes are unsuitable for the
existence of any observers. The lifetime $10^{-43}$ seconds is shorter
than the  lifetime of any virtual particle, so one may argue that
non-inflationary universes with ${1\over 2} \dot\phi^2 + {1\over 2}
(\partial_i\phi)^2 > V(\phi)$ do not really exist at the classical
level. Meanwhile all universes with ${1\over 2} \dot\phi^2 + {1\over
2} (\partial_i\phi)^2 \lesssim V(\phi)$ exist for an exponentially
long time and become exponentially large.  It seems therefore that
chaotic inflation occurs under rather natural initial conditions, if
it can begin at $V(\phi) \sim 1$ \cite{Chaot,book}. If one discards
universes with a total lifetime smaller than $10^{-40}$ seconds, just
as one discards subcritical bubbles during tunneling, then one may
argue that most of the universes with a macroscopically large lifetime
are inflationary.

Similar conclusions can be obtained if one considers the probability
of quantum creation of the universe from nothing.  According to
\cite{Creation}, the probability of quantum creation of the universe
is suppressed by $e^{-S}$, where $S= {3 \over 8 V(\phi)}$ is the
entropy of de Sitter space. This implies that creation of a closed
universe is most probable at $V(\phi) \sim 1$, and it is exponentially
suppressed for $V(\phi) \ll 1$. For example, the typical energy
density during inflation in new inflation scenario is $V(\phi) \sim
M^4_{\rm GUT} \sim 10^{-12}$; quantum creation of the universe in such
theories is suppressed by $e^{-10^{12}}$. Meanwhile in the simplest
versions of chaotic inflation with $V(\phi) \sim \phi^n$ inflation is
possible for $V(\phi)$ close to the Planck density as well as below
it, and the creation of inflationary universes is not exponentially
suppressed. A similar result can be obtained by a combinatorial
analysis proposed in \cite{Linde:1994wt}.

A simple way to understand this argument is to consider the usual
uncertainty relation $\Delta E\Delta t \sim 1$. The total energy of
matter in a closed inflationary universe is proportional to the volume
of the ``throat'' of de Sitter hyperboloid $H^{-3} \sim M_p^3
V^{-3/2}$ multiplied by the energy density $V$. This gives $\Delta E
\sim M_p^3 V^{-1/2}$. For $V \sim M_p^4$ one finds $\Delta E \sim M_p$,
so there is no problem, according to the uncertainty relation $\Delta
E\Delta t \sim 1$, with creating a Planck size closed inflationary
universe with $V \sim M_p^4$ during the Planck time $\Delta t \sim
M_p^{-1}$.

We are going to return to this issue in a separate publication.  The
main reason why we discussed it here was to present a general
description of initial conditions for chaotic inflation. As we see,
inflation may easily occur in a domain of a smallest possible size
$O(M_p^{-1}) = O(1)$. The universe initially may have a total mass as
small as $M_p = 1 \sim 10^{-5}$ g and it may contain no elementary
particles at all.  The expansion of this domain gives rise to a
domain of size $l \sim \exp{2\pi \phi_p^2} \sim \exp{2\pi m^{-2}} \sim
10^{10^{12}}$ \cite{Chaot,book}.  The decay of the scalar field at the
end of inflation \cite{DL,KLS94} creates about $10^{10^{12}}$ elementary particles; we
see only a minor part of them ($\sim 10^{88}$ particles) in the
observable part of the universe.

Moreover, once inflation begins in an interval $m^{-1/2} < \phi_0 <
m^{-1}$ in the theory $m^2\phi^2/2$, the universe enters an eternal
process of self-reproduction \cite{Eternal,LLM}. Thus, if inflation
begins in a single domain of the smallest possible size $l = O(1)$, it
makes the universe locally homogeneous and produces infinitely many
inflationary domains of exponentially large size. This makes the whole
issue of initial conditions nearly irrelevant: Non-inflationary
domains die within Planckian time, i.e. effectively remain unborn,
whereas inflationary domains exist for a long time and produce
infinitely many other inflationary domains.

\subsection{Comparison with the argument by Hollands and Wald}\label{Comp}

Now we should try to compare this picture with the argument by
Hollands and Wald suggesting that the initial conditions for inflation
cannot be natural. They formulated their argument in the following
form:

{\it Let $\mathcal U$ be the collection of universes that start from a
``big bang'' type of singularity, expand to a large size and
recollapse to a ``big crunch'' type of singularity. Let $\mathcal I$
denote the space of initial data for such universes and let
$\mu_{\mathcal I}$ denote the measure on this space used by the
``blindfolded Creator''. Let $\mathcal F$ denote the space of final
data of the universes in $\mathcal U$, and let $\mu_{\mathcal F}$
denote the measure on $\mathcal F$ obtained from $\mu_{\mathcal I}$
via the ``time reversal'' map. Suppose that $\mu_{\mathcal I}$ is such
that dynamical evolution from $\mathcal I$ to $\mathcal F$ is measure
preserving (``Liouville's theorem''). Then the probability that a
universe in $\mathcal U$ gets large by undergoing an era of inflation
is equal to the probability that a universe in $\mathcal U$ will
undergo an era of ``deflation'' when it recollapses.}  Then the
authors noted that the probability that a universe dominated by
ordinary matter will deflate is very small. From their perspective,
this implies, by time reversal invariance, that the probability of
inflation must also be very small.

As we see, this argument is completely unrelated to the investigation
of initial conditions for chaotic inflation performed in the previous
section. It is based on several formal assumptions about the evolution
of the universe that Hollands and Wald consider natural.   First of
all, they consider a recollapsing universe.  Such a recollapse will
not occur in our universe if we have a small cosmological constant
leading to the present stage of acceleration. However, this is a minor
problem since one can consider a formal time reversal at any moment of
time.

The most important assumption is that dynamical evolution is measure
preserving, which, roughly speaking, means that the number of degrees
of freedom does not change during inflationary evolution. This
assumption often holds for dynamical systems  
ignoring particle production.  However, in application to inflationary
cosmology this assumption is definitely incorrect. Indeed, in
inflationary cosmology the total energy and entropy of the scalar field and
particles created by its decay is not conserved. In our simple
scenario the total initial mass of matter in the universe was $O(1)
\sim 10^{-5}$ g, but later on its total energy became exponentially
large. In terms of particle physics, measure preservation implies
conservation of the number of particles. But in chaotic inflation
initially we did not have any particles at all, the total entropy was
$O(1)$, and then we got more than $10^{10^{12}}$ particles with
entropy greater than $10^{10^{12}}$. The absence of adiabaticity is a
key feature of all inflationary models because inflation removes all
particles that could be present before inflation; all $10^{88}$
particles that we see now within our cosmological horizon were created by the decaying scalar field.

Decay of the scalar field and particle production are irreversible
processes, and therefore, quite independently of the issue of
probabilities, time reversal of inflationary evolution {\it can never
produce the same initial conditions the universe started with.} The
scalar field that decayed at the end of inflation is not going to
re-appear again if one reverses the time evolution. The number of
particles produced by this field, just as the inhomogeneities produced
during inflation, will only grow on the way back to the singularity.

Initially our universe could have had Planck size and Planck mass at
the moment when it had Planck density. But then it became
exponentially large and heavy. Time-reversal of its evolution would
lead to an exponentially large and exponentially heavy universe at the
moment when it reached Planck density on its way back to the
singularity. Indeed, at the Planck density each Planck size volume can
contain no more than one particle with Planckian temperature. This
means that if the observable part of the universe (which is many
orders of magnitude smaller than the whole inflationary universe at
present) were to recollapse, it would consist of more than $10^{88}$
Planck size domains at the Planck density. And the whole universe
containing $10^{10^{12}}$ particles (even if we ignore
self-reproduction that makes this number indefinitely large) would
consist of $10^{10^{12}}$ Planck size domains at the Planck density.
{\it We cannot squeeze the universe back to its initial Planckian size
by time reversal}.

Similarly, large scale cosmological inhomogeneities created by quantum fluctuations are not going to disappear under time reversal. Even if the inflationary
universe was ideally homogeneous from the very beginning, time
reversal of its evolution from the present stage would never return it
to its original form.

Thus, inflationary evolution is irreversible, and the `obvious'
requirement of measure preserving evolution is not satisfied in
inflationary cosmology. Information contained in the single Planck
size initial inflationary domain is insufficient to predict the speed
and position of each of the $10^{10^{12}}$ particles created after
inflation. And even if we were able to know exactly the final
conditions after inflation including the speed and position of each of
the $10^{10^{12}}$ particles created after inflation, this would not
help us to squeeze all of these particles into the initial domain of
the Planck size.   This irreversibility comes about because the elementary particles and even galaxies that we see now appeared as a result of random quantum processes that could not be predicted by imposing initial conditions on the evolution of the classical scalar field $\phi$ in the initial  Planck size domain. This invalidates the
argument of Hollands and Wald against inflation.

These issues are deeply related to the discussion of reversibility in
quantum cosmology contained in the famous paper by Bryce DeWitt
\cite{Dewitt:yk} (see also the well-known paper by Hawking
\cite{Hawking:1985af} and comments by Don Page
\cite{Page:1985ei}). All equations of general relativity are time
reversal invariant, and therefore there are an equal number of
universes with growing and decreasing entropy. However, once we pick
up a large classical universe and define time there, the entropy and
the total number of particles inside this universe can only
grow. Similarly, once we consider a particular realization of an
inflationary universe after it becomes large and the scalar field
decays producing many particles, these particles cannot be un-born on
the way back to the singularity.

But even if we ignore quantum effects and particle production, and
even if it were possible to recreate the initial inflationary domain
by time reversal, this would not mean that investigation of the final
conditions after inflation can tell us anything about the probability
of initial conditions for inflation. Indeed, the inflationary regime
with $p \approx -\rho$ is an attractor in the phase space
$(\phi,\dot\phi)$ during expansion of the universe. Meanwhile during
the stage of collapse, the deflationary trajectory with $p \approx
-\rho$ is repulsive, and all trajectories are attracted to the regime
with the stiff equation of state $p \approx +\rho$
\cite{negpot}. Therefore the fact that the collapsing universe
typically does not deflate is completely compatible with the
possibility that the natural initial conditions in the early universe
lead to inflation. We will consider a particular example illustrating
this general statement in the next section.

\subsection{Inflationary regime as an attractor}

In this section we will discuss one of the aspects of the theory of
initial conditions for inflation. For illustrative purposes, we will
consider the simplest model $m^2\phi^2/2$ and study the evolution of
the homogeneous field $\phi$ in a flat universe filled with radiation.
Our goal here is not to study the most general initial conditions for inflation in a flat universe, but to illustrate on a simple example some basic issues related to the choice of initial conditions versus final conditions. For convenience, in this section we measure time in units of $m^{-1}$.

\hskip 1cm\FIGURE[!h]
{\centerline{\epsfig{file=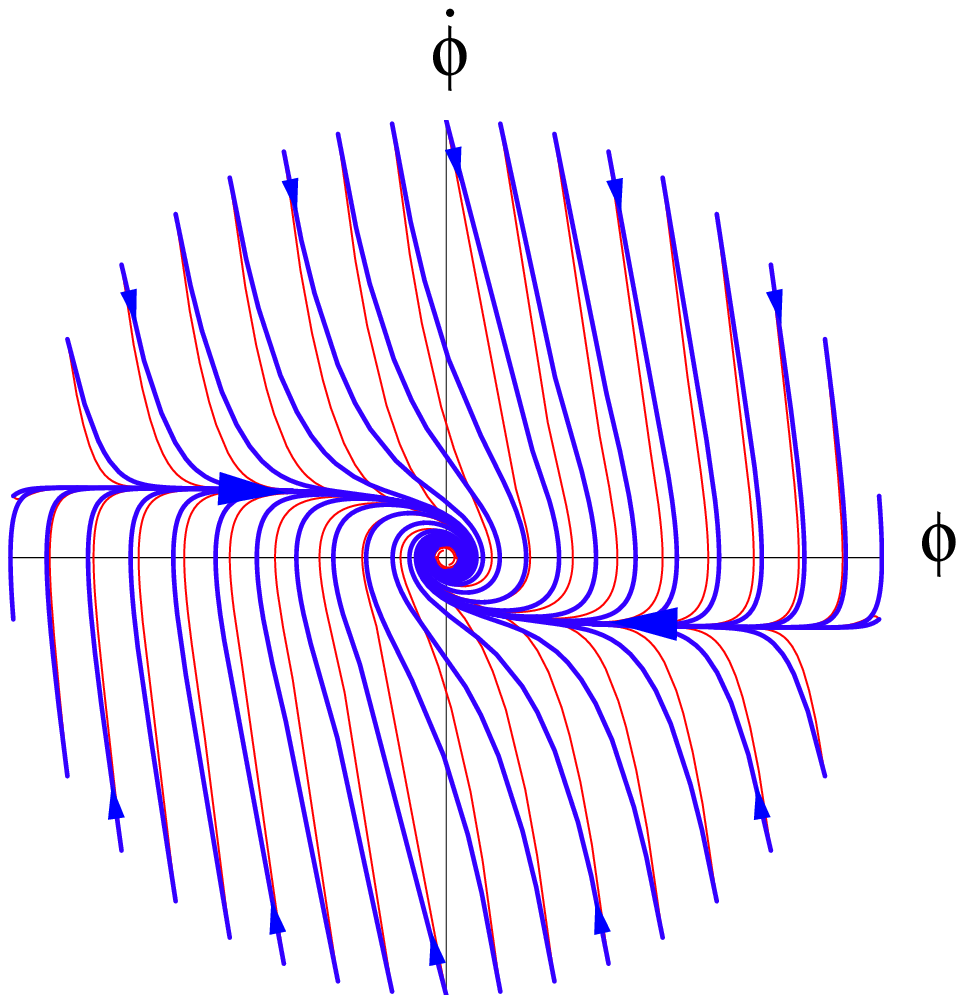,width=.55\columnwidth}}

\\

\caption{Phase portrait for the theory $V(\phi)={1 \over 2}m^2\phi^2$.
The blue (thick) lines show trajectories describing the universe
without radiation. The scalar field has half Planck density at the
beginning of the simulations. The red (thin) lines show trajectories
where an equal amount of energy in radiation was added to the
system. In both cases the velocity of the scalar field rapidly
decreases, which usually leads to the onset of inflation. The
inflationary separatrices (the horizontal lines to which most of the trajectories converge) are attractors when we move forwards in time, but they will be  repulsors if we reverse direction of time and move from the center back to large $|\phi|$ and $|\dot\phi|$.}
\label{comparison}}
  
The phase portrait of this model in the space of the variables
$(\phi,\dot\phi)$ is shown in Fig \ref{comparison} \cite{negpot}.  The
red (thin) lines in this figure show the evolution of
$(\phi,\dot\phi)$ in a universe filled with radiation. Evolution
begins at the Planck density sphere ${m^2\over 2} (\dot\phi^2 +\phi^2)
+\rho_{\rm rad} = 1$, where $\rho_{\rm rad}$ is the energy density of
radiation. (Recall that we are using time in units of $m^{-1}$.) The
plot describes the situation where initially the energy was equally
distributed between the scalar field and radiation. For comparison we
also show the trajectories of the scalar field in the absence of
radiation (thick blue lines).

As we see, most of the trajectories starting at the Planck density
approach the inflationary attractors $\dot\phi = \pm {1\over 2
\sqrt{3\pi}}$. In the Figure we have shown the process for the
unrealistically large mass $m = 1$. The situation becomes much more
impressive in the realistic case $m \sim 10^{-6}$. Indeed, one can
show that {\it the fraction of trajectories not approaching inflation
is smaller than $O(m^{-1})$ } \cite{Khalat,negpot}.

The simplest way to see this is to consider a regime where in the
beginning of the process one has $\rho_{\rm rad} \sim {m^2\over 2}
(\dot\phi^2 + \phi^2)$ for $\phi \gg 1$, see red (thin) lines in
Fig. \ref{comparison}. The kinetic energy of the scalar field
$\dot\phi^2/2$ in this regime decreases as $a^{-6}$. Meanwhile, the
density of radiation decreases as $a^{-4}$. Therefore the energy
density of radiation eventually becomes greater than
$\dot\phi^2/2$. As we will see, once this occurs the field rapidly
slows down or even completely freezes. This provides good initial
conditions for a subsequent stage of inflation \cite{Khalat,negpot}.

Consider the moment $t_0$ when the energy density of radiation becomes
greater than $\dot\phi^2/2$. In this regime (and neglecting $V(\phi)$)
one can show that $\dot\phi = \dot\phi_0 {a_0^3\over a^3} = \dot\phi_0
\left({t_0\over t}\right)^{3/2}$.  Even if this regime continues for
an indefinitely long time, the total change of the field $\phi$ during
this time remains quite limited.  Indeed,
\begin{equation} \label{changephi}
\Delta\phi \leq \int\limits_{t_0}^\infty \dot\phi dt = \dot\phi_0
\int\limits_{t_0}^\infty \left({t_0\over t}\right)^{3\over 2}
dt = 2~\dot\phi_0 t_0 .
\end{equation}
If $t_0$ is the very beginning of radiation domination, then one has
$m^2\dot\phi_0^2/2 \sim \rho_{\rm total}/2 = {3 H_0^2\over
8\pi}$. This implies that $H_0 \sim mt_0^{-1} \sim 3
m\dot\phi_0$. Therefore
\begin{equation}\label{rchange}
\Delta\phi < 1   
\end{equation}
in Planck units (i.e. $\Delta\phi < M_p$).   

This simple result has several important implications.  In particular,
if the motion of the field in a radiation-dominated universe begins at
$|\phi| \gg 1$, then it can move only by $\Delta\phi \lesssim
1$. Therefore in theories with flat potentials the field always
remains frozen at $|\phi| \gg 1$.  It begins moving again only when
the Hubble constant decreases and $|3H\dot\phi|$ becomes comparable to
$|V,_\phi|$. But in this case the condition $ 3H\dot\phi \approx
|V,_\phi|$ automatically leads to inflation in such theories as
$m^2\phi^2/2$ for $\phi \gg 1$.  This means that all trajectories
starting at $|\phi|>1$ enter a stage of inflation. Since the field
$\phi$ initially can take any value in the interval from $-m^{-1}$ to
$m^{-1}$, and inflation does not happen only for $|\phi| \lesssim 1$,
the fraction of non-inflationary trajectories is $O(m^{-1})$
\cite{Khalat,negpot}. For $m = 10^{-6}$ this means that out of a
million trajectories equally distributed over the initial Planckian
sphere ${m^2\over 2}(\dot\phi^2 + \phi^2) = 1$ only one trajectory
will be non-inflationary!

One can represent the  position of the vector $(\phi,\dot\phi)$ at the Planckian
sphere ${m^2\over 2}(\dot\phi^2 + \phi^2) = 1$ by introducing the  angle $\theta_i$ such that initially $\phi ={\sqrt{2}\over m} \cos \theta_i$, $\dot\phi= {\sqrt{2}\over m}  \sin \theta_i$.
Then the assumption of equal distribution over the sphere ${m^2\over 2}(\dot\phi^2 + \phi^2) = 1$ implies that all values of $\theta_i$ are equally probable. In this case, as we have seen, the probability to have inflation is $1-O(m) \approx 1$.

Let us see, however, what will happen if we impose a similar condition after the end of inflation. When the field $\phi$ becomes much smaller than $1$,
inflation ends. At this stage $\phi$ begins oscillating near the
minimum of the effective potential at $\phi = 0$ and the vector
$(\phi,\dot\phi)$ begins rotating around the origin in the phase space
(i.e. around the point $(\phi=0,\dot\phi=0)$) with a slowly decreasing
amplitude. After a long time, the decrease of the amplitude during
each oscillation becomes extremely small, so the phase portrait
becomes similar to the phase portrait of a simple harmonic oscillator,
$(\phi=\phi_f \cos (t+\theta_f),\dot\phi= \phi_f \sin (t+\theta_f))$),
where  $\theta_f$ depends on the initial values of
$(\phi,\dot\phi)$ at the Planck time, i.e. on $\theta_i$.  According to our analysis, almost all trajectories in the interval $-\pi <\theta_i<\pi$ will approach the inflationary separatrix,  
Fig. \ref{comparison}, and therefore almost all  trajectories after inflation  will have the same value of $\theta_f$. There will be some trajectories with different
values of $\theta_f$, but the ``density'' of such trajectories will be
suppressed by the factor $O(m^{-1}) \sim 10^{-6}$.
 
Now let us make a time-reversal and consider the state of this
harmonic oscillator as an initial state for the trajectories going
back to the singularity.  The initial state for this process is
described by $\phi_f$ and $\theta_f$. {\it A priori}, no initial phase
$\theta_f$ is any better than any other phase. Therefore the natural
probability measure for initial conditions in the time-reversed
universe should not depend on $\theta_f$.  But we know that only a
fraction of $O(m^{-1}) \sim 10^{-6}$ out of the total range of the
values of $\theta_f$ in the interval from $0$ to $2\pi$ correspond to
inflationary trajectories.

Thus, from the point of view of the natural measure on space of
initial conditions (equal density of trajectories crossing the Planck sphere
${m^2\over 2} (\dot\phi^2 +\phi^2) +\rho_{\rm rad} = 1$) the
probability to have inflation is $1- O(m^{-1}) = 1-O(10^{-6})$.
Meanwhile, if one studies the same process from the point of view of
the time-reversal process and uses the natural probability measure on
space of final conditions, as suggested by Hollands and Wald, one
finds that the probability of inflation is $10^{-6}$. But this trick
in fact says nothing about the probability of inflation; it just a
reflection of the well-known fact that even though dynamical equations
are time-symmetric, the initial and final conditions are not
interchangeable.  The fact that the inflationary separatrix is a
repulsor and the probability of deflation is $10^{-6}$ when we move
back in time does not have any implications for the probability of
inflation when we move forward in time. Indeed, when we move forward
in time, the inflationary separatrix is an attractor, see
Fig. \ref{comparison}, and the probability of inflation in this model
is $1- O(m^{-1}) = 1-O(10^{-6})$.

If one considers the possibility of an inhomogeneous distribution of
the scalar field, the probability of inflation becomes somewhat
smaller but still remains large in the model under consideration, just
like in the closed universe case considered in Section \ref{cl}, see
\cite{Chaot,book,LLM}.

Note, that in this section we considered the simplest, intuitive choice of measure on the space of initial conditions for $\phi$ and $\dot\phi$, following \cite{Khalat}. Holland and Wald assumed that there should exist a canonical measure preserved during the dynamical evolution in accordance with the Liouville's theorem.  Indeed, a canonical measure in space of all dynamical variables $a$, $\dot a$, $\phi$, $\dot\phi$ does exist \cite{Henn,Gibb,HawkPage}. One could expect that by using this measure one can evaluate the probability of inflation in the early universe by considering various conditions at the late stages of the evolution of the universe  and then 
going back in time, as suggested by Hollands and Wald. 

However, it turns out that the use of the canonical measure \cite{Henn,Gibb,HawkPage} does not imply that the results of calculation of the probability of inflation are invariant  along the trajectories.  The integrals involving this measure   diverge, which makes the results ambiguous and depending on the cosmic time when the probability is evaluated \cite{HawkPage}. According to \cite{HawkPage}, if one investigates the probability of inflation using this measure  starting at very low energy density, the probability of inflation could seem very small. However, if one imposes initial conditions at large energy density (i.e. in the very early universe), 
the probability of inflation appears very large. In particular, if one imposes initial conditions at the Planck time, which is the most reasonable choice because only after that moment one can describe our universe in terms of classical space-time,   one finds $P\sim 1- O(m^{-1})$  \cite{HawkPage}, 
in agreement with the results of Ref. \cite{Khalat} and with the 
estimates obtained in our paper.

This means that our simple intuitive approach based on the measure in terms of the angles $\theta$
in the space of variables $\phi$ and $\dot\phi$ leads to the same qualitative results as the  approach
based on the canonical measure \cite{Henn,Gibb,HawkPage}.

Finally, we should remember again that the simple investigation performed in this section, as well as the investigation using the canonical  measure in the space of variables $a$, $\dot a$, $\phi$, $\dot\phi$ \cite{HawkPage}, ignores the issue of quantum effects, which make the evolution of the universe completely irreversible, see Section \ref{Comp}.  There is no way one could store the information about the $10^{10^{12}}$ particles produced after inflation by fixing appropriate initial conditions in a single domain of initial size $O(1)$ in Planck units, and it is impossible to squeeze all of these particles back to the initial inflationary domain. There is no one-to-one correspondence between the present state of the universe and the initial conditions for the scale factor and the classical scalar field $\phi$ at the beginning of inflation. As a result, one can say almost nothing about the  initial conditions at the beginning of inflation by considering the time-reversal of the present evolution of the universe.

\section{Conclusions}

In this paper we analysed the argument of Hollands and Wald suggesting
that inflation does not solve any of the major cosmological
problems. Their argument was based on the observation that if our
universe were to collapse back to the singularity, it would not
deflate. Therefore they argued that it could not inflate on its way to
its present state. We do not think that this argument is valid. The
inflationary regime is an attractor for solutions for the scalar field
during  expansion, but it is a repulsor for the solutions during
contraction of the universe. Moreover, the dynamics of inflation are
completely irreversible due to particle production after inflation and
creation of inhomogeneities during inflation. Therefore the
investigation of the time-reversed behaviour of a typical post-inflationary universe tells us almost nothing about the initial conditions that produced the
universe. Meanwhile, an investigation performed here and in
\cite{Chaot,book,Creation,Linde:1994wt,Eternal,LLM,Khalat} suggests
that initial conditions for inflation in the simplest versions of
chaotic inflation are quite natural, and inflation does indeed solve
the major cosmological problems.

We also showed that the new mechanism of generation of density
perturbations proposed by Hollands and Wald is very problematic. In
particular, in their scenario the inhomogeneities responsible for the
large scale structure observed today were generated at an epoch when
the energy density of the hot universe was $10^{95}$ times greater
than the Planck density. This makes all predictions concerning such
density perturbations completely unreliable. We have shown that the
only way to avoid this problem is to assume that there was a stage of
inflation in the early universe.

The authors are grateful to S.~ Hollands, R.~Wald and D. Page for illuminating
discussions and to G. Felder for his assistance. The work by L.K. was
supported by NSERC and CIAR. The work by A.L. was supported by NSF
grant PHY-9870115, and by the Templeton Foundation grant
No. 938-COS273.  L.K. and A.L. were also supported by NATO Linkage
Grant 97538. V. M. is grateful to Princeton University for the
hospitality.

\end{document}